\documentclass{article}

\usepackage{arxiv}

\usepackage[utf8]{inputenc} 
\usepackage[T1]{fontenc}    
\usepackage{hyperref}       
\usepackage{url}            
\usepackage{booktabs}       
\usepackage{amsfonts}       
\usepackage{nicefrac}       
\usepackage{microtype}      
\usepackage{lipsum}
\usepackage{graphicx}
\graphicspath{ {./images/} }

\title{Adaptive optics with reflected light and deep neural networks}

\author{
 Ivan Vishniakou \\
  Center of Advanced European\\ Studies and Research (caesar)\\
  53175 Bonn, Germany
   \And
 Johannes D. Seelig\thanks{Corresponding author: johannes.seelig@caesar.de} \\
  Center of Advanced European\\ Studies and Research (caesar)\\
  53175 Bonn, Germany\\
}

\begin{document}
\maketitle

\begin{abstract}
Light scattering and aberrations limit optical microscopy in biological tissue, which motivates the development of adaptive optics techniques.
Here, we develop a method for adaptive optics with reflected light and deep neural networks compatible with an epi-detection configuration. 
Large datasets of sample aberrations which consist of excitation and detection path aberrations as well as the corresponding reflected focus images are generated. These datasets are used for training deep neural networks.  
After training, these  networks can disentangle and independently correct excitation and detection aberrations based on reflected light images recorded from scattering samples. A similar deep learning approach is also demonstrated with scattering guide stars. The predicted aberration corrections are validated using two photon imaging. 
\end{abstract}


\section{Introduction}

Diffraction limited imaging in many samples, for example in biological tissue, is limited due to aberrations and scattering. In this situation, combining laser scanning microscopy with adaptive optics can improve image resolution by correcting for aberrations using wavefront shaping  \cite{kerr2008imaging, rodriguez2018adaptive, rotter2017light, booth2007adaptive}. 
Using reflected light for wavefront shaping has the advantage that it relies on intrinsic sample contrast \cite{dunn1996sources} and is therefore independent of sample labeling \cite{feierabend2004coherence, rueckel2006adaptive, kerr2008imaging}. Additionally, even for fluorescently labeled samples, particularly for dim or sparsely labeled samples, the scattered light signal can be larger than the fluorescent signal \cite{feierabend2004coherence, rueckel2006adaptive, booth2011adaptive}.  

Imaging with reflected light comes however with the difficulty that excitation and detection aberrations are not easily separated \cite{booth2011adaptive, kang2017high, yoon2019laser, badon2019distortion}.
In reflection-mode imaging, the signal depends on the excitation point spread function (PSF), which undergoes aberrations, the scattering object, and the detection PSF, which can undergo different aberrations \cite{kang2017high, yoon2019laser, yoon2020deep}. 

For making an accurate correction of the excitation focus under these conditions, therefore contributions to aberrations accumulated in the excitation and reflected detection pathway through the sample need to be disentangled. For strongly scattering samples, separation of illumination and detection pathways has been achieved using matrix methods \cite{yoon2020deep, kang2017high, yoon2019laser, badon2019distortion}, which however require a large number of phase sensitive measurements to obtain a correction.
For weakly scattering samples, one option to separate excitation and detection aberrations is to use a wavefront sensor that only detects light that is reflected from a tightly constrained focal volume. This can be achieved using coherence gating with high-bandwidth light sources combined with interferometric detection \cite{feierabend2004coherence, rueckel2006adaptive}. 
A technically less demanding approach was implemented using confocal imaging with a pinhole sufficiently large to retain wavefront information \cite{rahman2013direct}. In this situation, however, due to the extended confocal volume, aberrations measured with a wavefront sensor depended on a combination of sample and detection characteristics \cite{rahman2013direct, booth2011adaptive, booth2007adaptive}. 

More recently, approaches for wavefront sensing based on deep neural networks have been developed \cite{angel1990adaptive, sandler1991use, paine2018machine, swanson2018wavefront, andersen2019neural, zhang2019machine, ma2019numerical, jin2018machine, nishizaki2019deep, hu2019learning, xu2019improved}. These neural network methods directly compare measured light distributions to computationally generated ones.
These approaches work so far only for a configuration that requires the correction of a single pass through a scattering medium, such such as the atmosphere in astronomy and therefore cannot be applied for reflected light detection. In the epi-detection configuration commonly used in biological imaging, aberrations in the excitation and detection pathways need to be considered.

Here, we extend the deep learning approach for transmission wavefront sensing to reflection-mode imaging in an epi-detection configuration. Reflected foci for different aberrating samples are generated by independently modulating the excitation as well as detection path of the microscope using a spatial light modulator. With this setup, imaging for example a reflecting planar object through an aberrating layer can be modeled by modulating both, the excitation as well as the detection pathway while at the same time recording the light reflected off a mirror at the sample plane. Imaging a guide star through an aberrating layer can be modeled by only modulating the detection pathway while forming a focus on the mirror with the unmodulated excitation pathway. 

Datasets generated in this way are used to train deep neural networks. We show that after training, these neural network models can disentangle excitation and detection aberrations of scattering samples. We verify the resulting excitation corrections using two-photon imaging. 

\section{Results}
The experimental approach is shown schematically in Fig.~\ref{fig:scheme}. A two-photon microscope is combined with a spatial light modulator (SLM) and reflected light detection for wavefront sensing and correction. The reflected focal spot is monitored at three different focal planes using cameras (see for example \cite{angel1990adaptive, hanser2003phase, ma2019numerical}). Different from other reflection-mode adaptive optics approaches \cite{booth2002adaptive, feierabend2004coherence, rueckel2006adaptive, rahman2013direct} we separate excitation and detection such that both pathways can be modulated independently. 

This is motivated by that fact that excitation and detection aberrations can differ. Generally, the incoming beam undergoes aberrations which alters the PSF at the focal plane. This secondary light source again undergoes sample aberrations on the return path. Due to the extended nature of the secondary light source, the return path does not necessarily overlap with the excitation path, thus resulting in different aberrations for the two pathways. This is illustrated with a simulation in Fig.~\ref{fig:simulation_results} (see Methods for details on simulation). Fig.~\ref{fig:simulation_results} a shows the simulated optical system, a simplified version of the actual setup: light is focused through an aberrating phase mask onto a reflecting surface with a lens. The reflected light passes again through the lens and the phase mask and is focused onto a detector with a second lens.  Fig.~\ref{fig:simulation_results} b shows the resulting reflected focus image (bottom row) after passing a plane wave (top row) through the optical system (see figure legend for details). As seen in Fig.~\ref{fig:simulation_results} c, for a weak aberration consisting of radial even Zernike modes, the combined aberrations accumulated in the excitation and detection pathway double (as expected, see \cite{booth2011adaptive}), resulting in the interference pattern in the focal volume (bottom row). In contrast, odd radial Zernike modes cancel \cite{booth2011adaptive}, and a diffraction limited focal volume is preserved (Fig.~\ref{fig:simulation_results} d, bottom row). However, for stronger aberrations from odd radial Zernike modes (Fig.~\ref{fig:simulation_results} e) differences in the excitation and detection pathway lead to only incomplete cancellation of aberrations in the focal volume. Differences in excitation and detection path lead to the asymmetric reflected phase pattern (Fig.~\ref{fig:simulation_results} e, second row from top) and result in a distorted reflected focal volume. This also suggests that in such a situation the odd components can be detected.

Taking such differences in excitation and detection aberrations into account, we generated datasets modeling three different sample configurations.  First, we modeled the situation of a planar reflecting object with entirely uncorrelated excitation and detection aberrations with randomly selected Zernike modes of up to order 28 in both pathways (Network~1, see Methods for details).  Secondly, modeling the expected similarity between excitation and detection  aberrations in weakly scattering samples, we used a dataset with a variance of $\pm30$ \% between excitation and detection Zernike coefficients to constrain the sample space (Network~2). 
Third, to model scattering from a guide star, we generated a dataset with only detection modulations, while focusing with the unmodulated excitation beam at the mirror surface (Network~3).

These datasets of (pairs of) phase modulations together with the corresponding reflected focus images,  were then used for training deep neural network models. 
The pairs of phase patterns in the excitation and detection path served as an output, and images of the reflected focal spots from the three cameras in different focal planes served as the input. After training, these networks predict Zernike coefficients for the excitation as well as the detection pathway based on images of the focal light distribution obtained when imaging through scattering samples. In this way, a correction of the excitation beam that is not confounded by aberrations in the reflected beam can be extracted.

\subsection{Methods}

\subsubsection{Setup}

\begin{figure}
    \centering
    \includegraphics[width=0.67\textwidth]{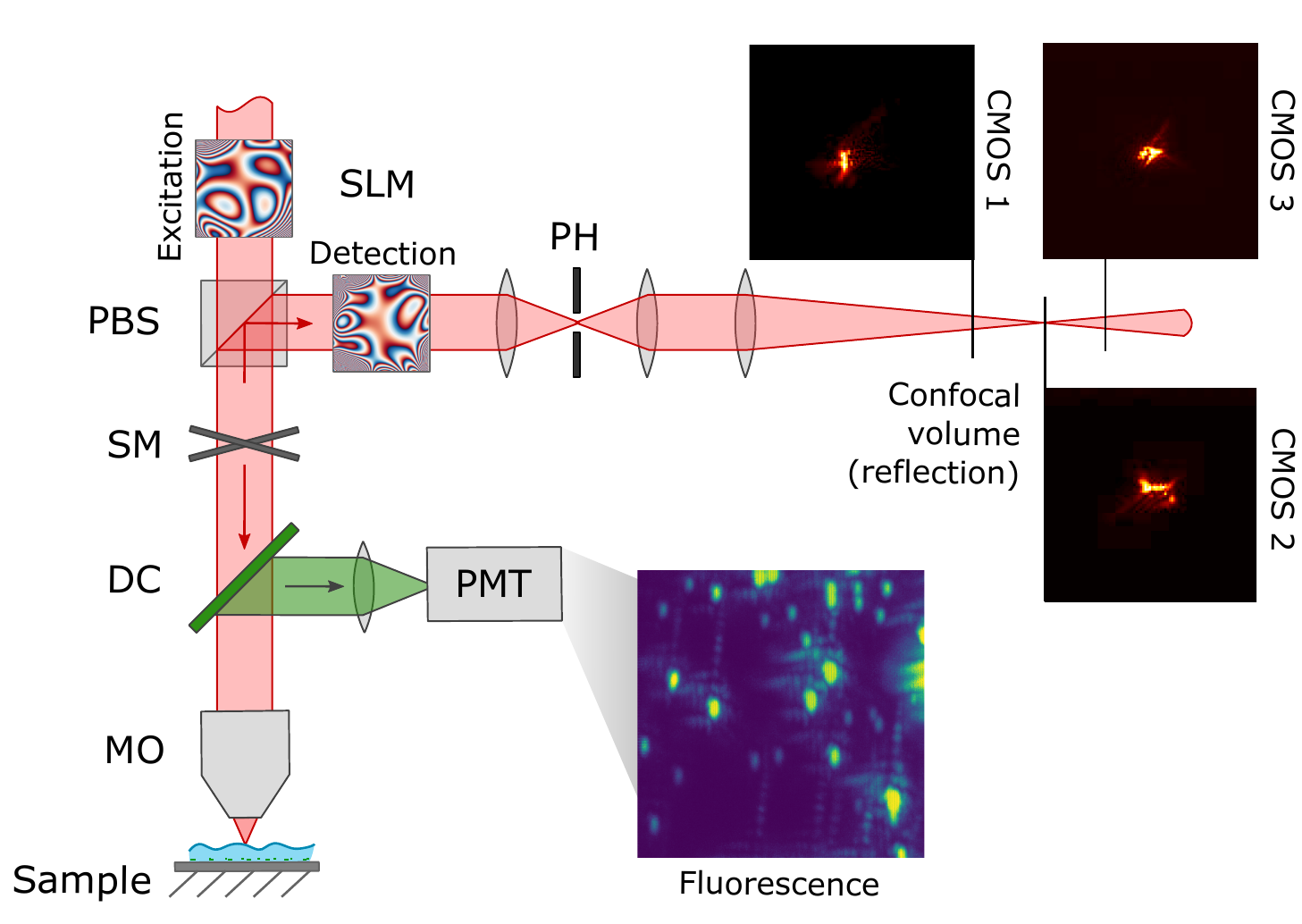}
    \caption{Setup schematic. Fluorescence is observed using two-photon scanning microscopy. Excitation and detection pathways are each controlled with an SLM. Reflected light is imaged in an epi-detection configuration onto three cameras. MO = microscope objective, DC = dichroic mirror, SM = scanning mirrors, PBS = polarizing beam splitter, SLM Excitation = spatial light modulator in excitation pathway, SLM Detection = SLM in detection pathway, PH = pinhole, CMOS 1-3 = cameras, PMT = photomultiplier tube. See legend to Fig.~\ref{fig:setup_schematic_SI_final} for details.}
    \label{fig:scheme}
\end{figure}

\begin{figure}
    \centering
    \includegraphics[width=1\textwidth,trim={0 0 2cm 0},clip]{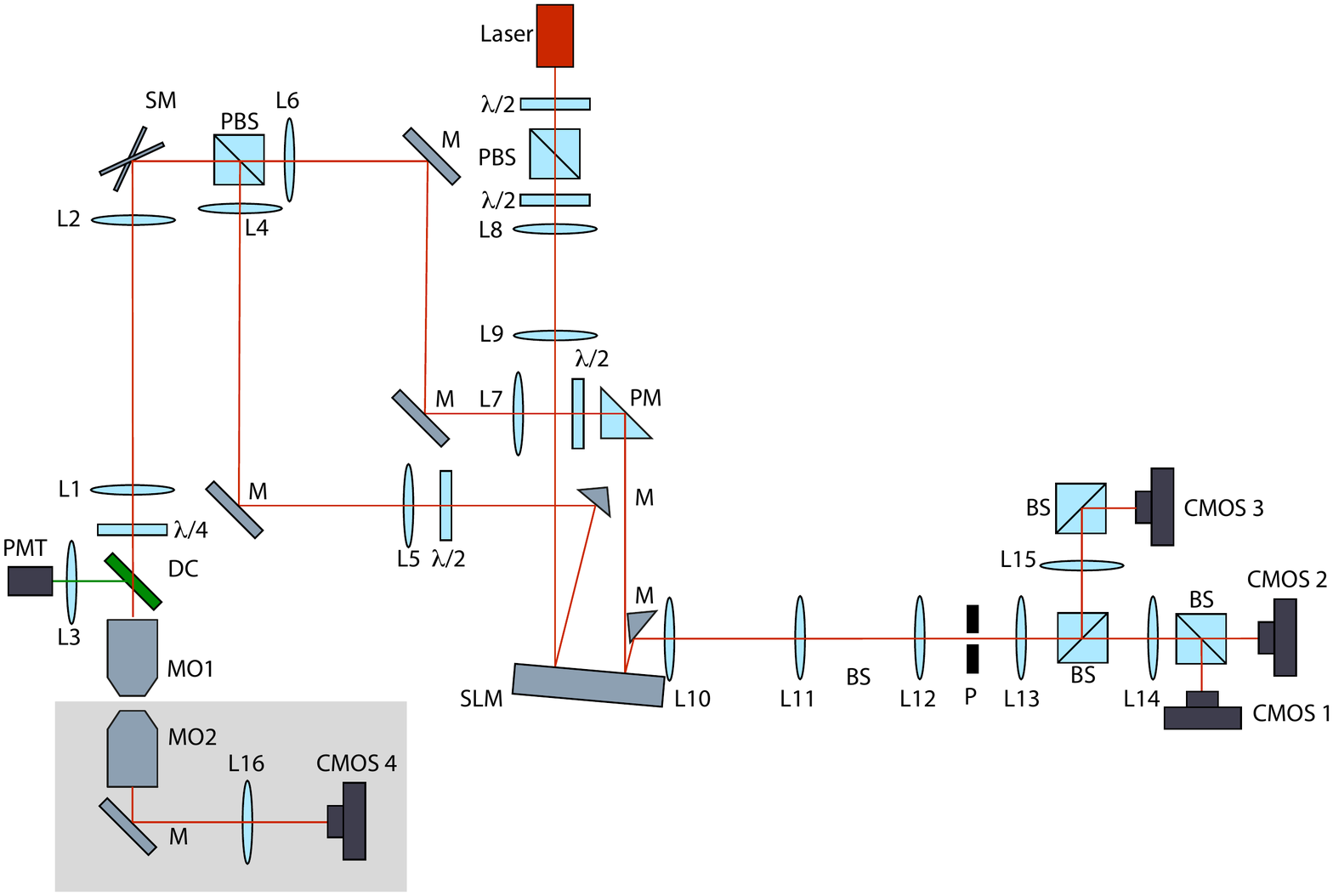}
    
    \caption{MO1 = Nikon 16 x, N.A. 0.8, water immersion objective (CFI75 LWD 16X W),  MO2 = Olympus 40x, N.A. 0.8, LUMPLFLN. All lenses were from Thorlabs, achromatic doublets, antireflection coated for 650 to 1050 nm. Focal lengths were (in mm) L1 = 300, L2 = 30, L3 = 50, L4 = 125, L5 = 300, L6 = 100, L7 = 400, L8 = 75, L9 = 250, L10 = 150, L11 = 150, L12 = 40, L13 = 50, L14 = 75, L15 = 75, L16 = 300. PBS = polarizing beam splitter, BS = beam splitter, (both anitreflection coated for 650 nm to 1050 nm), $\lambda/2$ = polymer zero order half-wave plate (WPH05ME-980), PM = reflecting prism mirror, P = pinhole with 300 $\mu$m diameter, CMOS = CMOS cameras (Basler, acA640-750um), SLM = spatial light modulator (Meadowlark, HSP1920-1064-HSP8), M = mirror, DC = dichroic mirror, SM = resonant scanning mirror, PMT = photomultiplier tube for fluorescence detection (two-photon imaging). The transmission pathway (light-gray box) was only used for observing the resulting corrections for a stationary excitation beam and was not used to compute corrections.}
    \label{fig:setup_schematic_SI_final}
\end{figure}

The setup, shown schematically in Fig.~\ref{fig:scheme} and in detail in Fig.~\ref{fig:setup_schematic_SI_final}, consists of a custom-built two-photon microscope equipped with a resonant scanner and controlled through ScanImage \cite{pologruto2003scanimage} with an added detection path for reflected light. Both the excitation path, as well as the reflected detection path are independently modulated with a spatial light modulator (SLM). 

The excitation beam is expanded, reflected off the SLM, demagnified, and imaged onto the scanner through a polarizing beam splitter (PBS) (see legend of Fig.~\ref{fig:setup_schematic_SI_final} for details). The (linear) polarization direction  is adjusted for maximum transmission through the PBS with a half-wave plate.  The scanner is imaged onto the back focal plane of the objective. A quarter wave plate is placed after the tube lens to achieve circular polarization and to optimize reflected light transmission through the PBS with orthogonal linear polarization with respect to the excitation light \cite{rueckel2006adaptive}.  Reflected light is detected in a descanned configuration through the polarizing beam splitter, imaged onto a different part of the SLM and imaged onto a pinhole. The pinhole is imaged onto three cameras in different focal planes using 50/50 beam splitters and relay lenses. One camera focal plane was selected at the focus, one in front and one behind the focus (see for example \cite{angel1990adaptive, hanser2003phase, ma2019numerical}).

For independent excitation and detection path modulation, the SLM was divided into two equal parts, each having a size of $960\times1080$ pixels. The SLM was controlled through custom software written in Python using Blink SDK provided by Meadowlark Optics.
To precisely center the phase modulations displayed on the SLM with respect to the beam, for initial alignment a center-symmetric phase pattern was displayed subsequently in both parts of the SLM.  The center pixel of the respective SLM window was found by moving the pattern until the reflected focus was center-symmetric. The polarization direction of the beam imaged onto the SLM was optimized for modulation using lambda-half plates in both the excitation and detection pathways (see Fig.~\ref{fig:setup_schematic_SI_final}).

\subsubsection{Relationship between excitation and detection path phase patterns}

\begin{figure}
    \centering
    \includegraphics[width=0.55\textwidth,trim={0 0 0 0},clip]{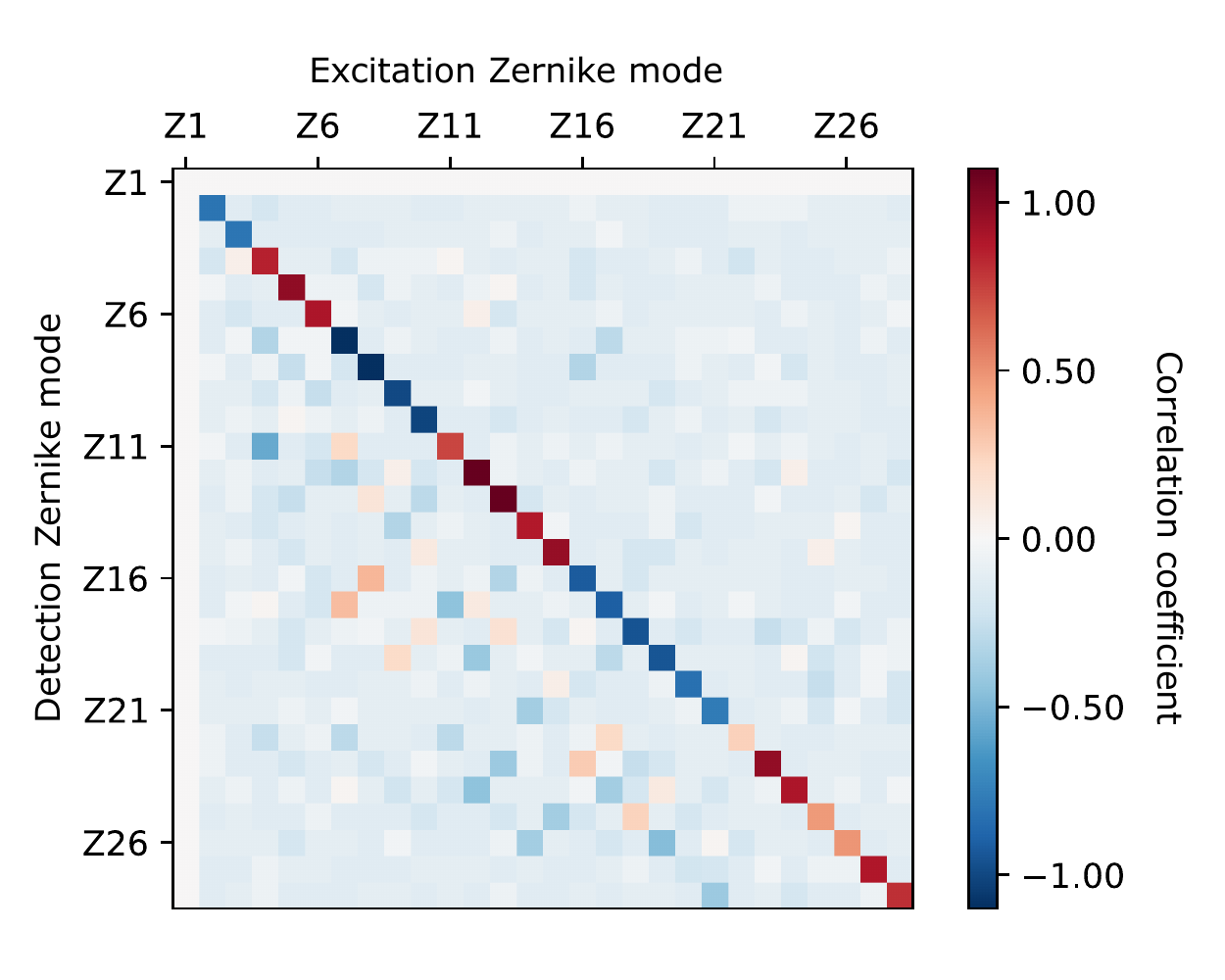}
    \caption{Linear correspondence model between Zernike modes displayed in the excitation and detection pathway (see Methods for details). Even components in the excitation are sensed with the same magnitude in the detection, and odd ones are sensed with a negative magnitude. }
    \label{fig:correlation_matrix}
\end{figure}

To calibrate the relation between excitation and detection pathways, which was required for experiments with guide stars (see below), we calculated a linear correspondence model (matrix) between excitation and detection Zernike modes. For this purpose a mirror was placed at the sample plane and random modulations were displayed in the excitation path. The resulting reflected confocal images were fed into Network 3 (trained on detection modulations only). With a set of 10 000 modulations applied in the excitation path and predicted as detection-path aberrations, a matrix relationship was determined (Fig.~\ref{fig:correlation_matrix}). The magnitude difference between excitation and detection phase corrections in the two pathways was taken into account in all generated datasets. The matrix shows that, as expected, even components in the excitation are sensed with the same magnitude in the detection, and odd ones are sensed with a negative magnitude.

\subsubsection{Simulations}
\begin{figure}
    \centering
    \includegraphics[width=1\textwidth,trim={0 0 0 0},clip]{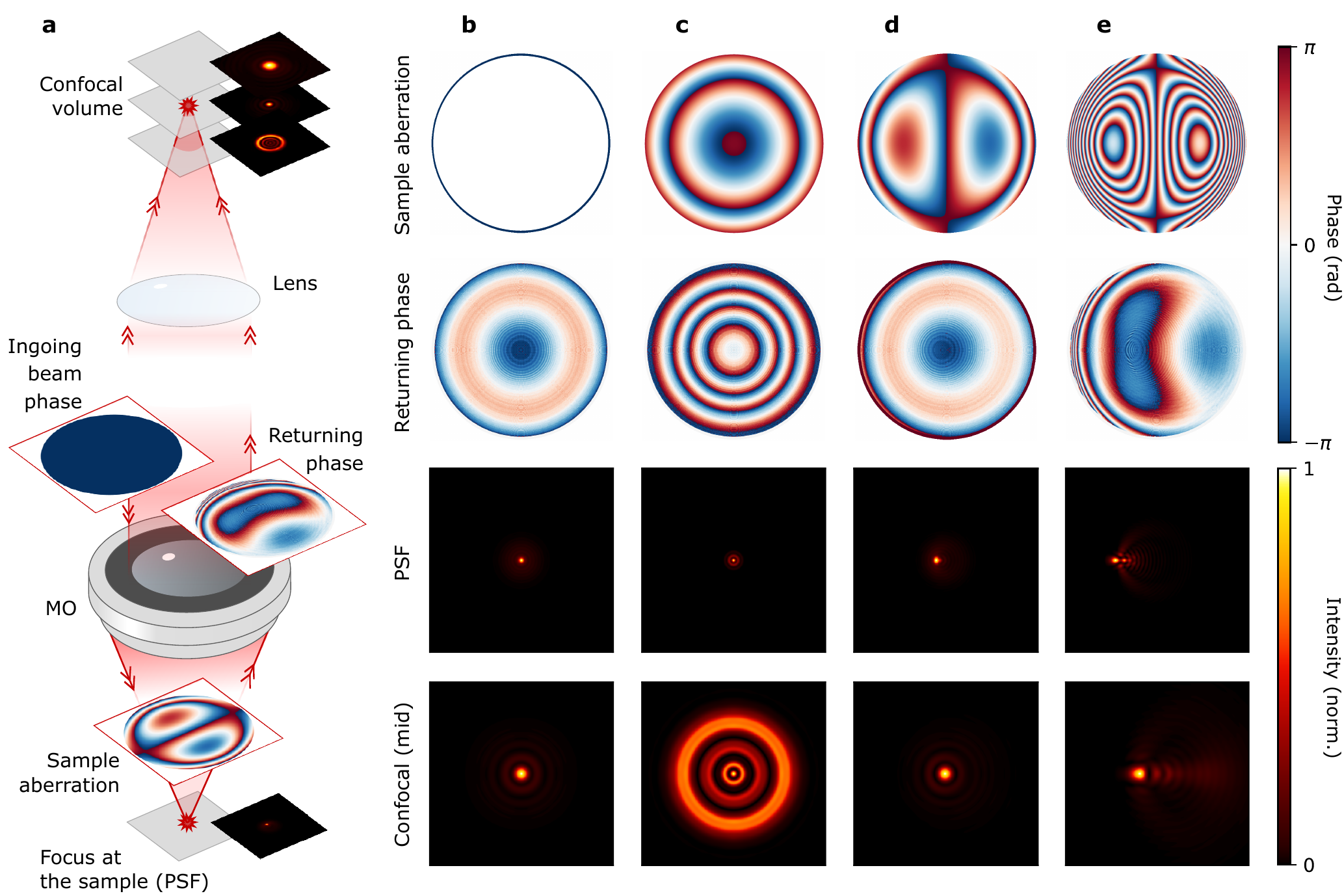}
    \caption{Simulation of impact of different aberrations on resulting reflected focal volume. \textbf{a} Schematic of simulated optical system (see Methods for details). A flat wavefront enters the microscope objective and undergoes sample-induced phase modulation at the distance of 0.5 mm form a mirror surface on the excitation and reflected return path; the returning beam is focused with a 200 mm lens. The PSF at the sample and reflected focal planes are simulated. \textbf{b} Example with no aberrations. Top row: flat wavefront, second row: reflected wavefront, third row: focus at sample plane, bottom row: focus at reflected volume.  \textbf{c} Aberration of even radial order which magnifies itself on the return path. \textbf{d} Self-correcting aberration of odd radial order. \textbf{e} Same as \textbf{d}, but with higher magnitude of phase aberration, demonstrating the failure of self-correction in this situation.}
    \label{fig:simulation_results}
\end{figure}

Using a Rayleigh-Sommerfeld solver \cite{diffractio}, we simulated light propagation through the microscope and sample and monitored the point spread function at the sample and reflected focal plane (Fig.~\ref{fig:simulation_results}). The simulated pathway is illustrated in Fig.~\ref{fig:simulation_results}~a: a flat wavefront enters the microscope objective, propagates to the surface of an aberrating layer (which adds a spatial phase modulation to the beam), reaches the mirror (the PSF is calculated in this plane), is reflected back, passes again through the lens and the aberrating layer and is finally focused with a lens in the reflected focal plane (see legend of Fig.~\ref{fig:simulation_results} for parameters used).\\

\subsubsection{Data processing and neural network approach}

The CNN architecture consisted of a cascade of 4 convolutional layers with ReLu activations (64 filters $11\times11$ with stride $4\times4$ and batch normalization, 64 filters $5\times5$ with stride $2\times2$ and batch normalization, 128 filters $3\times3$ with $2\times2$ max pooling, and 192 filters $3\times3$ with $2\times2$ max pooling respectively). These layers are followed by a dense layer with 3072 elements, sigmoid activation and 0.3 dropout regularization, and an output dense layer with linear activation, the size of which corresponds to the number of predicted Zernike modes. The networks were trained by minimizing mean absolute error (MAE) of the prediction with Adam optimizer at learning rate 0.0001.

The confocal volume (imaged in 3 planes) was normalized by dividing by $255$ to bring the 8 bpp image to the $0\dots1$ range and was stacked into a $192\times192\times3$ tensor serving as input for the network. The output was the corresponding phase modulation, represented as vector of Zernike coefficients $(Z1\dots Z28)$; in cases where both excitation and detection modulations were used, they were both concatenated into as single vector $(Z1_{\textrm{exc}}\dots Z28_{\textrm{exc}}, Z1_{\textrm{det}}\dots Z28_{\textrm{det}})$. Each random modulation was generated by shuffling a harmonic sequence $1.5\pi/n$ and randomly choosing the sign of each of its elements. For both excitation and detection the Z1 (piston) mode was set to 0.

\subsection{Disentangling excitation and detection phase modulations}

We first tested whether neural networks could extract random, independent phase modulations displayed in the excitation and at the same time in the detection pathway based on the resulting reflected focus patterns. We therefore generated a dataset by displaying different random Zernike modes of up to order 28 simultaneously in each pathway. A mirror was placed at the sample plane of the microscope and the resulting reflected focus images were recorded. This models a planar sample with entirely uncorrelated excitation and detection aberrations. 

A network (Network 1) was trained on 180 000 such pairs of excitation and detection phase modulations and the resulting sets of confocal images.  Fig.~\ref{fig:networks_training}, top row, shows examples of predicted and target excitation and detection phase modulation with representative mean absolute errors. The mean absolute error (MAE) of the examples is also indicated in the full MAE distribution shown in the bottom row. The histogram of mean absolute error (MAE, Fig.~\ref{fig:networks_training}, bottom row) between prediction and target modulations is compared to the error for random pairing (gray). These results shows that the trained neural network can reliably disentangle and predict independent excitation and detection phase patterns based on reflected confocal images resulting from the combined modulation.

\subsection{Focusing through aberrating layers}

We next tested whether such networks could be used to separate excitation and detection aberrations based on reflected images from actual scattering samples and whether the resulting corrections could be used for focusing through the encountered aberrations. 
In these experiments, we focused through a layer of vacuum grease onto a reflecting surface. To be able to directly monitor the focus after the scattering layer, we focused on a fifty-fifty beam splitter. This resulted in a reflected focus image and at the same time allowed monitoring the focus at the sample surface with a transmission camera for a stationary beam (non-descanned transmission detection, see Fig.~\ref{fig:setup_schematic_SI_final}). To monitor the correction during scanning, we placed fluorescent beads (0.1 $\mu$m diameter) on the beam splitter surface and detected fluorescence using two-photon imaging.  

Corrections were computed using Network 2 which was trained on a dataset of 180 000 examples with a variance of $\pm30$ \% between excitation and detection Zernike coefficients. This allowed better coverage of the sample space compared to Network 1 (with similar sized training datasets) and was sufficient for correcting aberrations in weakly scattering samples. Examples of network predictions with representative MAEs are shown in Fig.~\ref{fig:networks_training}, top row. The MAE of the examples is indicated in the MAE distribution shown in Fig.~\ref{fig:networks_training}, bottom row.

\begin{figure}
    \centering
    \includegraphics[width=0.75\textwidth,trim={0 0 0 0},clip]{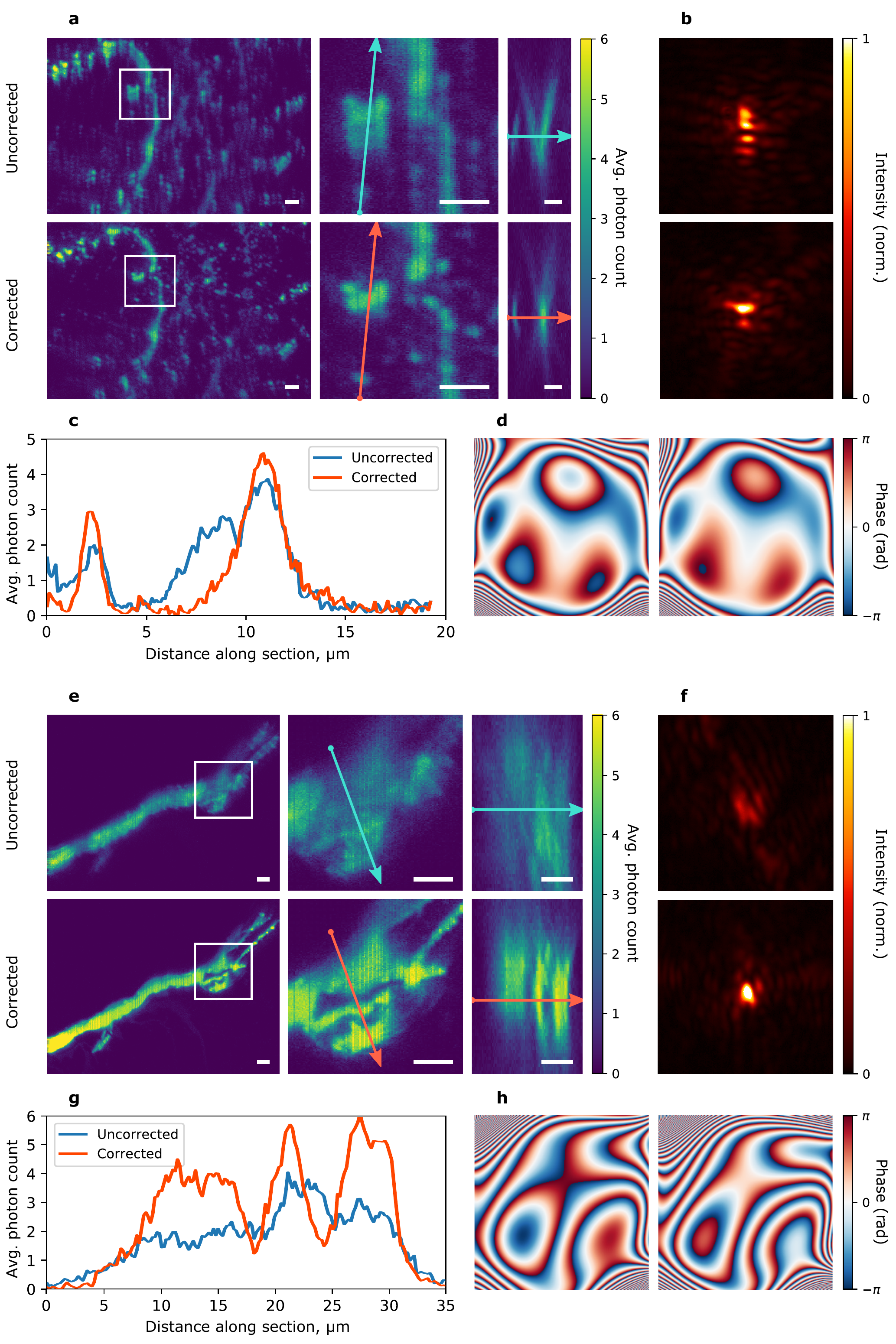}
    \caption{Aberration correction for imaging through scattering layer on planar reflector. \textbf{a} Left: example of corrected and uncorrected image of fluorescent beads distributed on reflector surface imaged through a layer of vacuum grease. Center: white frame in left figure. Right: axial cross section through lines in center figure recorded in a z-stack with 1 $\mu$ m step size. \textbf{b} Top: uncorrected focus at center of filed-of-view in \textbf{a}. Bottom: corrected based on reflected light. Colorscale is saturated in the corrected image, so that aberrations in the uncorrected image are visible. \textbf{c} Cross sections for uncorrected (blue) and corrected (red) images along the lines indicated in figure a, center. \textbf{d} Left: excitation and Right: detection phase mask. \textbf{e-h} Same as \textbf{a-d} for a second example. All scale bars are 5 $\mu$m.}
    \label{fig:corrections_mirror}
\end{figure}

Representative examples of two-photon images of fluorescent beads on the reflecting surface with and without excitation correction are shown in Fig.~\ref{fig:corrections_mirror}. Due to aberrations the center plane of the confocal volume is difficult to determine, and corrections were therefore obtained by averaging between three and five images of the confocal volume at different axial positions around the estimated center (separated by 2-5 micrometers). 
Confocal images were normalized the same way as training images (see Methods) and independently fed through the network and the predictions were averaged. 
The network output is vector of Zernike coefficients and excitation and detection aberration were generated. To correct for aberrations, the complex conjugate of the network output was displayed on the SLM.

Fig.~\ref{fig:corrections_mirror} a shows that aberrations in the fluorescence images can be corrected based on the reflected excitation light using the trained neural network. The improved focus (monitored with a transmission camera, \ref{fig:corrections_mirror} b and f), leads to improved resolution and signal as seen in \ref{fig:corrections_mirror} a and c, and e and g, respectively. (Note that the beads were placed on top of a beam splitter, so reflection from the beam splitter will likely distort the axial profile sown in Fig.~\ref{fig:corrections_mirror} a and e on the right side.) Overall, the decomposition into an excitation and detection pathway leads indeed to the formation of an improved focus suitable for two-photon imaging. 

We implemented a second imaging approach based on deep learning with scattering guide stars. Since for a guide star the observed focus distortion in the sample is a result of only return path aberrations, we trained a network on detection-only modulations (Network 3). Examples of predicted and target corrections are shown in Fig.~\ref{fig:networks_training}, top row. The MAE of the examples is shown in the distribution in the bottom row.

\begin{figure}
    \centering
    \includegraphics[width=0.8\textwidth,trim={0 0 0 0},clip]{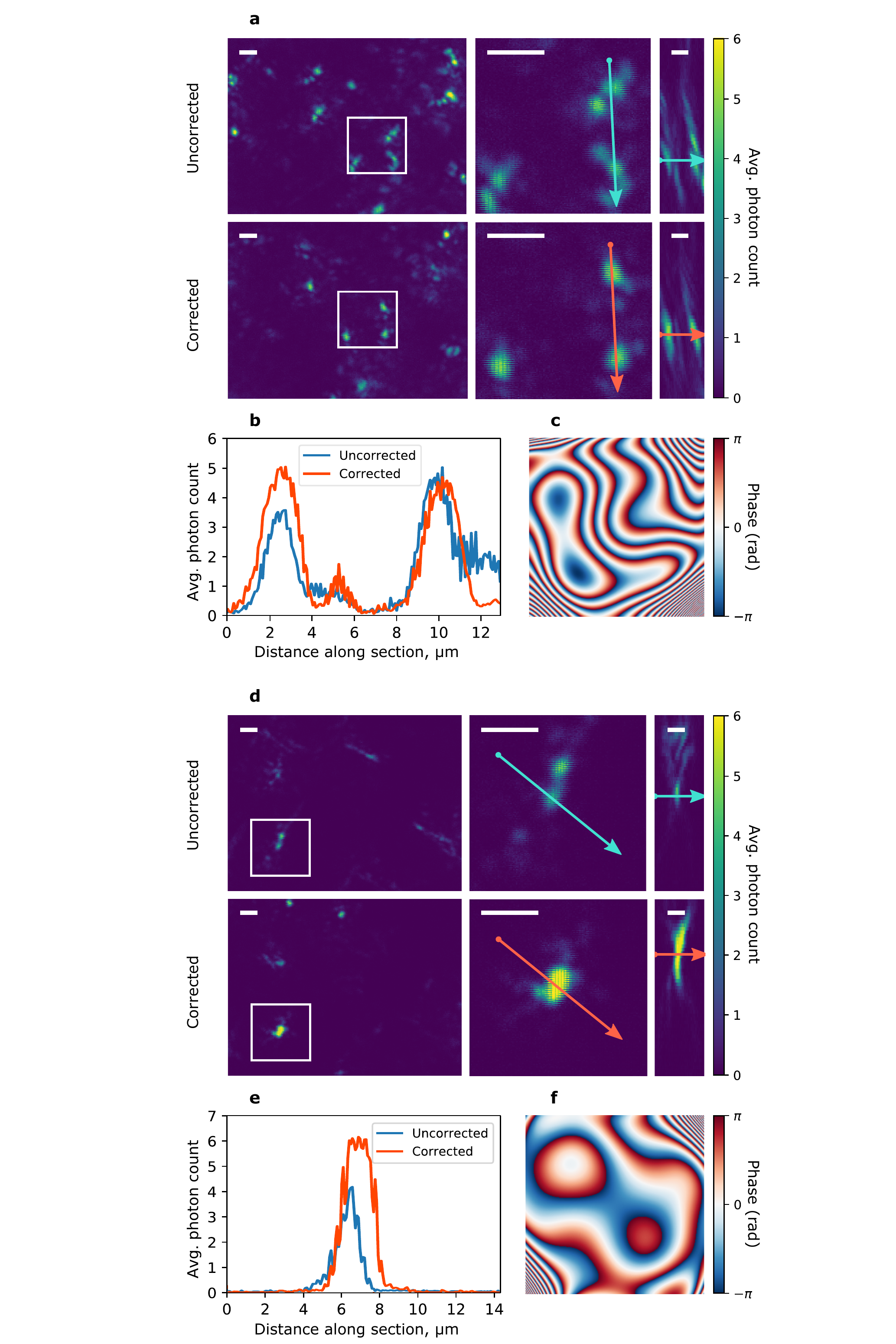}
    \caption{Aberration correction for imaging through scattering layer with guide stars. \textbf{a} Left: example of corrected and uncorrected image of fluorescent beads distributed in a volume of Agarose mixed with scattering guide stars imaged through a layer of vacuum grease. Center: white highlighted box in left figure. Right: axial cross section through lines in center figure recorded in a z-stack with 1 $\mu$ m step size. \textbf{b} Cross sections for uncorrected (blue) and corrected (red) images along the lines indicated in \textbf{a}, center. \textbf{c} Excitation phase mask. \textbf{d-f} Same as \textbf{a-c} for a second example. All scale bars are 5 $\mu$m.}
    \label{fig:corrections_guidestar}
\end{figure}

The confocal volume was imaged and corresponding aberration of the return path were predicted with the network. The correction is carried over to the excitation path using the correspondence model between excitation and detection path Zernike modes (see Methods and Fig.~\ref{fig:correlation_matrix}). The sample consisted of a mixture of scattering guide stars (9 $\mu$m diameter silver coated silica microspheres, Cospheric) and 0.1 $\mu$m fluorescent beads embedded in 1 $\%$ Agarose. (The transmission focus was not monitored directly in this case, since the focus was inside a volume, not at an interface as above.) The sample was again imaged through a layer of vacuum grease. Experiments were performed as above, with averaging over between three and five focal planes. Fig.~\ref{fig:corrections_guidestar} a and Fig.~\ref{fig:corrections_guidestar} d show two representative examples of corrections achieved with this approach. Both, resolution and intensity improve due to the applied corrections. 

\section{Discussion}
 
\begin{figure}
    \centering
    \includegraphics[width=1\textwidth,trim={0 0 0 0},clip]{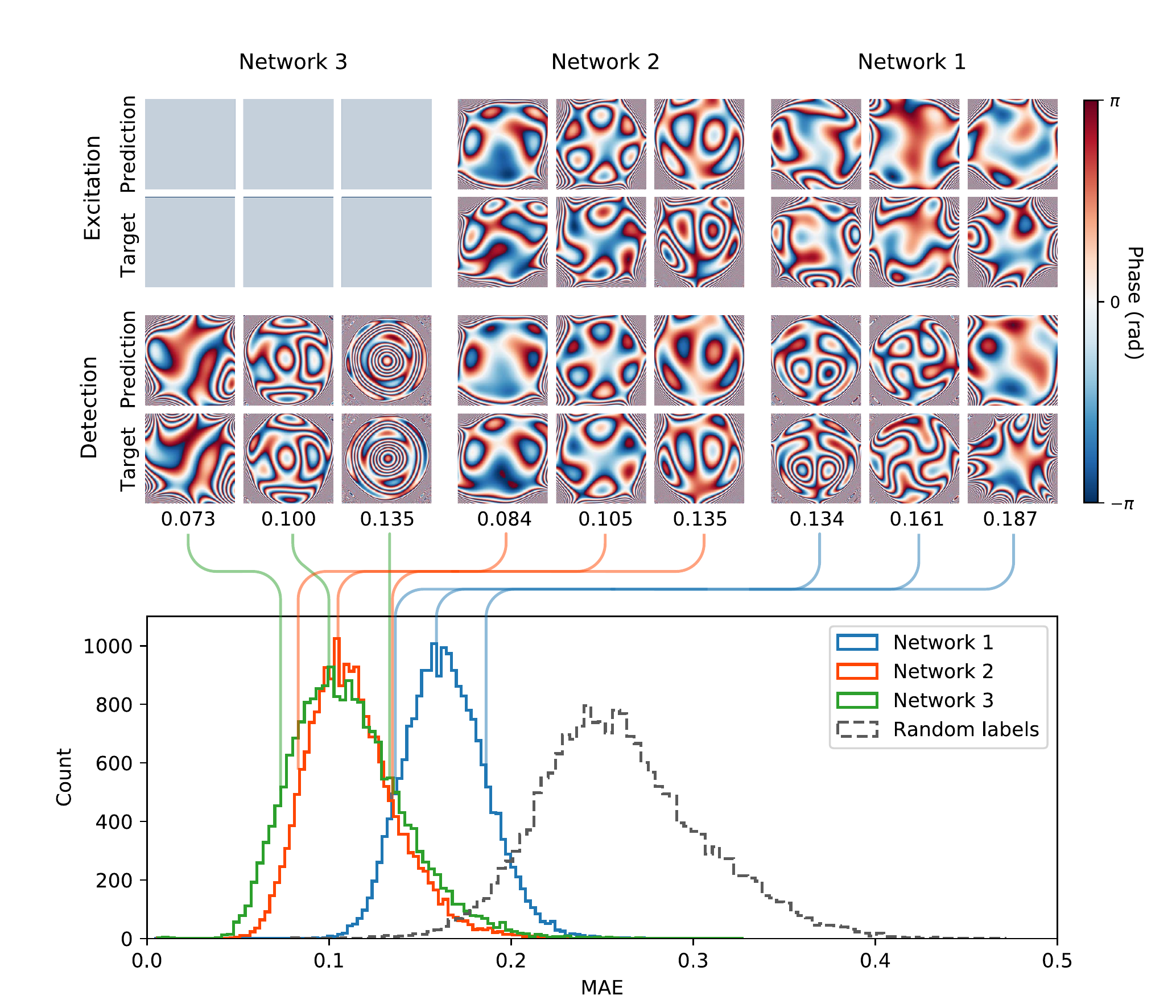}
    \caption{Top row: Examples of pairs of network predictions and targets for all three networks used for excitation pathway. Samples are drawn as indicated by colored lines at respective positions in the pertaining error distributions. Center row: Same as top row for detection pathway phase patterns. Bottom: Histograms of mean absolute error (MEA) between predicted and target phase masks for the three different networks, compared with errors obtained for random pairings.}
    \label{fig:networks_training}
\end{figure}

We have developed an adaptive optics approach for laser scanning microscopy based on reflected light imaging and deep neural networks. For network training large datasets of aberrated focus images were generated by combined excitation and detection pathway phase modulations, modeling the aberrations observed from extended reflecting objects in scattering samples. After training on such datasets, deep neural networks can extract underlying excitation and detection phase aberrations of modeled and actual samples based on reflected focus images (Fig.~\ref{fig:networks_training}). We validated this approach using two-photon imaging of fluorescent beads distributed on a mirror through an aberrating sample as well as by directly imaging the transmitted focus (Fig.~\ref{fig:corrections_mirror} and Fig.~\ref{fig:corrections_guidestar}). The resulting corrections achieved for a planar object rely neither on a tight focal volume nor a guide star, as typically  necessary for wave front sensing.

In a second approach, we obtained corrections by combining guide stars and deep neural networks. In this situation, training data was modeled using only detection path modulations and the corresponding corrections were then displayed in the excitation pathway for excitation correction. 

Reflected light detection generally has the advantage of being independent of sample labeling \cite{feierabend2004coherence}, but comes with the difficulty of separating excitation, detection, and sample contributions to the scattered signal \cite{yoon2019laser, yoon2020deep}. We addressed here the problem of separating excitation and detection aberrations with a deep neural network approach and independent phase modulation of both pathways. While we used either a reflecting surface or a guide star as reflectors, sample scattering characteristics could for example be included by recording training data from biological samples. 

For two-photon imaging experiments we introduced a coupling between excitation and detection corrections with a variance of $\pm30$ \% between excitation and detection Zernike coefficients to constrain the space of possible aberrations and to limit the required amount of training data. Using larger datasets will allow relaxing this constraint and to address larger differences in excitation and detection aberrations as observed in more strongly scattering samples \cite{yoon2019laser}. Additionally, we obtained best corrections by averaging over three to five different excitation focal planes and larger data sets will likely improve single shot predictions suitable for the fast frame rates required for dynamic biological samples.

Aberration corrections could additionally be improved by including higher order Zernike modes. In the current implementation, Zernike polynomials of up to order 28 were used, similar to previous approaches using wavefront sensors \cite{rueckel2006adaptive}. Up to 120 orders have for example been measured in a transmission configuration using a combination of deep neural networks and wavefront sensors \cite{hu2019learning}, suggesting that higher order modes could also be detected in reflection mode imaging. In addition to larger datasets, also different network architectures that would benefit from such larger datasets could be used, such as ResNet \cite{mockl2019accurate} or Inception \cite{andersen2019neural}.

Alternatively to Zernike polynomials (deep) neural networks can also be trained with different basis sets (see for example  \cite{turpin2018light}) which could potentially better match actual sample scattering characteristics. As an alternative to generating training data with an SLM, also entirely computationally generated datasets could be used (as for example in \cite{paine2018machine, andersen2019neural}). This would allow simulating datasets with scattered light distributions matched to the those observed in samples of interest.
Overall, the approach outlined here offers a versatile framework for excitation and detection aberration corrections that is independent of sample labeling and can be integrated with laser scanning microscopy.

\section*{Funding}
This work was supported by the Max Planck Society and the research center caesar. 
 
 
\section*{Disclosures}
The authors declare no conflicts of interest.

\bibliographystyle{unsrt}  
\bibliography{submission}  


\end{document}